\documentclass{article}
\usepackage{ijcai20}
\usepackage[english]{babel}
\usepackage{times}

\usepackage{soul}
\usepackage{url}
\usepackage[hidelinks]{hyperref}
\usepackage[utf8]{inputenc}
\usepackage[small]{caption}
\usepackage{graphicx}
\usepackage{amsmath}
\usepackage{booktabs}

\title{AI-based Monitoring \& Response System for Hospital Preparedness towards COVID-19 in Southeast Asia}

\author{
Tushar Goswamy$^1$\footnote{Contact Author}\and
Naishadh Parmar$^{1}$\and
Ayush Gupta$^1$\and
Raunak Shah$^1$\and
Vatsalya Tandon$^1$\and
Varun Goyal$^1$\and
Sanyog Gupta$^1$\and
Karishma Laud$^1$\and
Shivam Gupta$^1$\and
Sudhanshu Mishra$^1$\and
Ashutosh Modi$^1$\and
\affiliations
$^1$Indian Institute of Technology (IIT) Kanpur, Kanpur, India\\
\emails
\{tgoswamy, naishadh, ayushgup, raunaks, vatsalt, varung, sanyog, kslaud, guptash, sdhanshu, ashutoshm\}@iitk.ac.in
}

\begin{document}

\maketitle

\begin{abstract}
This research paper proposes a COVID-19 monitoring and response system to identify the surge in the volume of patients at hospitals and shortage of critical equipment like ventilators in South-east Asian countries, to understand the burden on health facilities. This can help authorities in these regions with resource planning measures to redirect resources to the regions identified by the model. Due to the lack of publicly available data on the influx of patients in hospitals, or the shortage of equipment, ICU units or hospital beds that regions in these countries might be facing, we leverage Twitter data for gleaning this information. The approach has yielded accurate results for states in India, and we are working on validating the model for the remaining countries so that it can serve as a reliable tool for authorities to monitor the burden on hospitals.       
\end{abstract}

\section{Introduction}

Social Media websites like Twitter and Facebook encourage frequent user expressions of their thoughts, opinions, and random details of their lives. India has the 8th largest user base of Twitter in the world, with 13.15 million users and growing, followed by Indonesia with 11.8 million users \cite{Statista}. This highlights the potential for gaining useful insights from the tweets posted by millions of users in these countries. Tweets and status updates range from significant events to inane comments. Most messages contain little informational value, but the aggregation of millions of messages can generate valuable knowledge. Twitter users often publicly express personal experience about overcrowding at hospitals, difficulties faced due to a shortage of equipment by them or their relatives and other issues arising due to the pandemic, which can help understand the ground reality of the situation. Previous research has studied the correlation between Twitter trends and influenza rates using tweets about the symptoms \cite{ICWSM112880}. Statistical techniques have been used to forecast flu rates using twitter data \cite{10.1371/journal.pcbi.1004513}. Influenza rates have been monitored at the local level in the USA during the Influenza Epidemic of 2012 \cite{10.1371/journal.pone.0083672}. Similarly, Signorini et al\cite{10.1371/journal.pone.0019467} have studied the correlation between twitter data and H1N1 cases for tracking of the infection. 

In this study, we are using the Twitter data of users to study the surge in hospitalization volumes due to the COVID-19 pandemic. We have focused our work on India, Indonesia, and Bangladesh for the scope of this study, with plans to extend this approach to other geographies in South-East Asia. Our research aims to identify incidents of overcrowding at hospitals, shortage of critical equipment like ventilators, and lack of available ICU units. This system can help understand the medical preparedness levels of the health facilities in these countries and the burden on their hospitals as the pandemic spreads. The system pipeline includes scraping historical tweets at a granular level to obtain a corpus, processing the corpus using Natural Language Processing tools, calculating signals from the processed data, and finally evaluating the results by comparing ground reports and bulletins. We have deployed Neural Translation models to account for the usage of regional language.

Our primary contribution to the AI community through this research is to demonstrate the application of an NLP-based Twitter model to monitor the burden on health facilities due to the COVID-19 pandemic. To the best of our knowledge, this is the first and the only approach of its kind, which can detect the trends in the worst-hit regions accurately based on Twitter data. We are closely working with members from WHO's Regional Office for South-East Asia (WHO-SEARO) to study and monitor our model's signals, and it is intended to help them with monitoring the situation in these countries and in identifying regions which are facing a resource crunch due to the pandemic. Our model can thus be used by public health organisations to recommend appropriate actions to the authorities in the regions which the model has identified.

\section{Data Extraction and Pre-processing}
\subsection{Natural Language Processing for Tweets}
\subsubsection{Historical Tweet Extraction}

We used the getOldTweets3 API \cite{GetOldTweets:2009} to scrape and extract historical tweets from the Twitter website. Unfortunately, Twitter has some restrictions due to which we are unable to
access all the tweets beyond seven days from the date of scraping. This leads to a misleading spike in the data (Fig. \ref{fig:full}). To address this, we scaled the tweets using the factor of change across
the peak.
\subsubsection{Data Cleaning}

To eliminate noise in the data and extract the important information, we performed the following
operations on the tweet corpus:
\begin{itemize}
    \item Removing Website Links: To prevent the same information from being captured twice.
    \item Removing non-ASCII characters: To eliminate noise and focus on relevant keywords only
    \item  Removing Stopwords: Removed words like `is', `an', `the' to focus on hospital-related words in the frequency analysis
    \item Tokenisation: We utilized the NLTK TweetTokenizer API \cite{Loper02nltk:the} to tokenize tweets. This was done to aid the keyword calculation process in subsequent steps.
    \item Lemmatisation: Implemented lemmatization on the tokens obtained for each tweet to convert the higher form of each word to their base forms.
\end{itemize}
We observed that the Indonesian tweets were heavily code-mixed as Indonesian Bahasa and English. Thus we implemented a modified version of the pipeline described by Barik et al. \cite{Barik2019NormalizationOI} to normalize and process the Indonesian tweets before calculating the scores.
For tweets from Bangladesh, the majority of the tweets were not codemixed and were either in the Roman English script or in the Bengali script. Thus, we processed the English tweets using the same set of operations mentioned above and implemented tokenization and normalization for the Bangla tweets.
\subsubsection{Keyword Selection}
To shortlist keywords which are most relevant to our analysis and can yield accurate signals for
the trend, we first created a corpus of common words related to the study like `hospital’, `ICU’, etc. This was followed by applying Topic Modelling using Latent Dirichlet Allocation \cite{10.5555/944919.944937}, to find words under similar category as our initial corpus. Topic Modelling provides clusters of similar words based on their usage, as well as their
weight to indicate how closely the words of a cluster are related. We also performed an n-gram analysis to find the frequency of these keywords in our corpus. This was followed by finding the most similar words to these keywords using Word2Vec \cite{mikolov2013efficient}. It allowed us to create vector representations for all the words in the vocabulary by taking into account the lexical as well as semantic features of the word. The context of all the keywords was studied to minimise noise in our corpus by avoiding irrelevant words/phrases, and at the same time ensuring that the critical signals are captured. Finally, based on the approaches outlined above, we shortlisted the following keywords for India, Indonesia and Bangladesh: 
\begin{itemize}
    \item India: `Hospital', `Medical College', `Beds', `ICU', `Shortage'
    \item Indonesia: `Hospital', `Medical', `Beds', `ICU', `Shortage', `Rumah Sakit', `Tempat Tidur', `Over kapasitas', `Penuh', `IGD', `UGD', `Fasilitas Kesehatan'
    \item Bangladesh: `Hospital', `Medical', `Beds', `ICU', `Shortage', `kendra'
\end{itemize}
\section{Twitter Score Calculation}
\subsection{Filtering}
We collected tweets since 1st March 2020, for all 29 states in India, 34 provinces in Indonesia and the country of Bangladesh as a whole, using the keywords: `corona', `covid' and `hospital'.

\subsection{Score Calculation}
We experimented with different combinations of scores for the model, and finally shortlisted the following based on the requirements of public health agencies who will use this model:

\subsubsection{Twitter Word Count/Day}
We obtain the Twitter Word Count/Day plot by calculating the daily count of the shortlisted keywords for a region. It is aimed at capturing incidents of overcrowding of hospitals as well as the shortage of beds and critical equipment.

\subsubsection{Twitter Volume/Day}
The Twitter Volume/Day score is calculated as the count of all the words in the filtered tweets. This indicates
the trend in the volume of tweets related to the COVID-19 pandemic in that region.

\section{Data Adjustment}

\subsection{Adjusting the peak}
\begin{figure}[!htb]
\begin{center}
  \includegraphics[width=\linewidth]{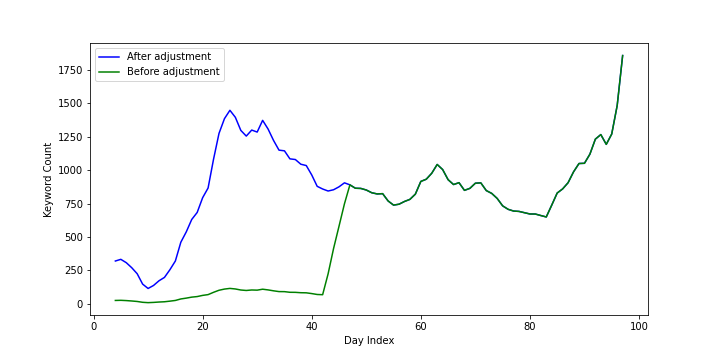}
  \caption{Twitter Word Count/Day plot before and after adjustment for Delhi}\label{fig:full}
\end{center}
\end{figure}

We discovered an abrupt
peak in both the plots mentioned in the previous section. After a
thorough analysis and observing the trend by re-scraping the data for a week, we found that the peak shifts by a day, if we scrape the data again, and always occurs at the 7th historical day from the date of scraping. This can be attributed to a possible restriction imposed by Twitter on accessing historical tweets. To overcome this issue, we normalised the historical tweets older than 7 days using the ratio of values across the peak. This was done since the full volume of tweets are scraped for the most recent 7 days, and the issue only arises for the tweets which are older than 7 days from out date of scraping.
The original and adjusted plots for Delhi can be seen in Fig. \ref{fig:full} 

\subsection{Data Modelling}

When we directly plot the data, it picks up the noise in the data, and this is visible as random fluctuations. This can be misleading in the analysis, and thus we ‘smooth’ the data by statistical techniques. We experimented with the following smoothing techniques and shortlisted the approach which gave the highest correlation with the positive cases data:

\begin{itemize}
    \item Moving Averages: We successively plot the average of $n$ -days (which is the window size) to get a smoother curve which captures the overall trend better. Different values of $n$ (3/4/5 etc.) are
tested, and the one which gave the highest correlation with the positive cases is selected \cite{MovingAverage:2009}
    \item Exponential Smoothing: we assign a higher weight to
the most recent observation when computing the moving averages. Under this approach, the
most recent observation should get a little more weight than the 2nd most recent, and the 2nd most recent should get a little more weight than the 3rd most recent, and so on \cite{MovingAverage:2009}
   \item Holt's Linear Exponential Smoothing: This technique includes two smoothing constants, one for the level and one for the trend.  Two equations, one for an estimate of the local level, and the local trend's estimate are applied iteratively to each point, that apply exponential smoothing \cite{MovingAverage:2009}.
\end{itemize}

We compared the Pearson Correlation Coefficient from the results of these techniques with the positive cases data and  found the 5-day Moving Average to give the highest correlation and thus, the best results.

\section{Results}

Since social media data is sensitive to political events, we marked the major political events of each country on the plots and studied the peaks which did not overlap with any major national events. We analyzed the trends for the worst-hit states and provinces, studied the tweets corresponding to the peaks, and compared them with news reports and bulletins to validate our results. A detailed analysis of Maharashtra (Fig. \ref{fig:maha}), Delhi(Fig. \ref{fig:del})(Worst-affected states in India) and Kerala(Fig. \ref{fig:ker}) (state in India where the cases have started falling, and it did not witness any overcrowding or shortage incidences at hospitals) has been provided below. For Indonesia and Bangladesh, we are monitoring the trends and fine-tuning the model to capture the signals accurately. These two countries' results have not been included in this paper as the work is still in progress.

\subsection{Maharashtra}
\begin{figure}[h]
    \centering
    \includegraphics[height=1.5in, width=8.0cm]{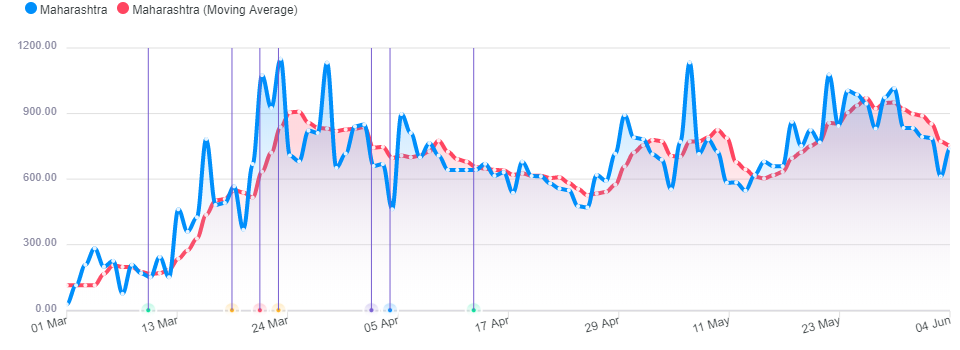}
    \caption{Twitter Word Count/Day Plot for Maharashtra, with major political events marked as vertical lines}
    \label{fig:maha}
\end{figure}

\begin{figure}[!htb]
\begin{center}
  \includegraphics[width=\linewidth]{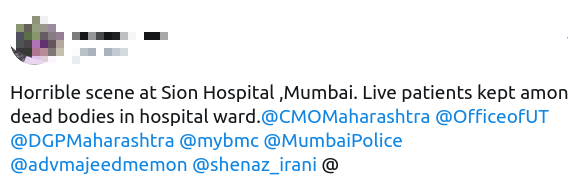}
  \caption{30th April Sample Tweet for Maharashtra}\label{fig:mum1}
  \includegraphics[width=\linewidth]{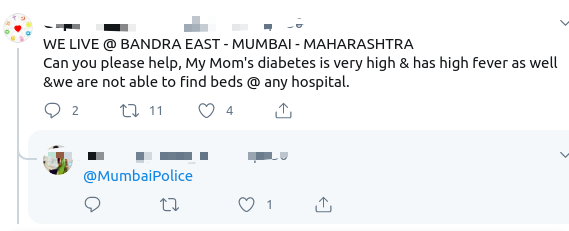}
  \caption{7th May Sample Tweet for Maharashtra}\label{fig:mum2}
\end{center}
\end{figure}

We observe major peaks near 30th April and 7th May, as seen in Fig. \ref{fig:maha} . We studied the tweets corresponding to these timestamps to understand the rise in the usage of the selected keywords like `hospital' and `overcrowding'. We found that majority of the tweets were indicative of the rise in hospitalisation numbers, as well as the increase in the incidences of overcrowding at hospitals in cities like Mumbai which is the financial capital of India and the most populated city of Maharashtra. Some sample tweets can be seen in Fig. \ref{fig:mum1} and \ref{fig:mum2}. We validated this information using official news reports about these incidents \cite{Darroch.2017}. The overall trend is also increasing and the Moving Average is at a higher level compared to March, which is in agreement with ground reports that the situation in hospitals is worse now compared to March \cite{mum2017}.

\subsection{Delhi}
\begin{figure}[h]
    \centering
    \includegraphics[height=1.5in, width=8.0cm]{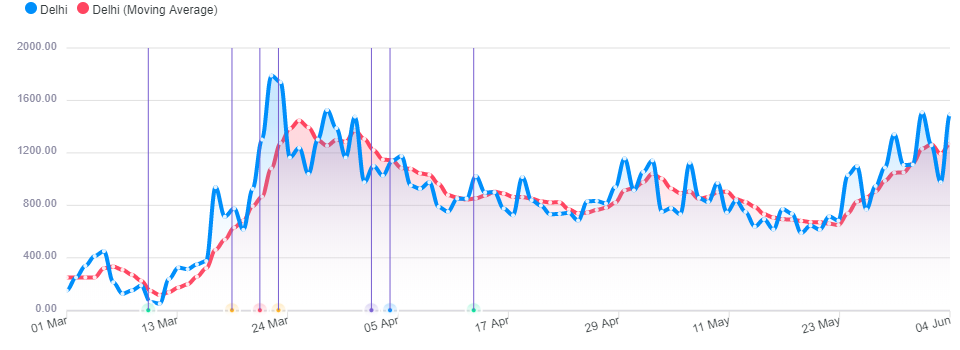}
    \caption{Twitter Word Count/Day Plot for Delhi}
    \label{fig:del}
\end{figure}

We observed peaks in Delhi at earlier dates compared to Maharashtra, which was verified by news reports confirming overcrowding and shortage of beds at major hospitals in Delhi like LNJP, Deen Dayal Hospital, etc. Peaks near 24th March, 14th April, 30th April, 25th May, 29th May, 1st June and a rising trend thereafter can be seen in Fig. \ref{fig:del}. Similar to Maharashtra, we found that most of the tweets corresponding to these peaks were indicative of the increasing burden on the health facilities in Delhi and the resource crunch faced by hospitals. The official news reports \cite{del1}, \cite{delhi} and \cite{delhilast} confirm the incidences reported by the tweets and observed as peaks on the plots. Also, the Moving Average is at a higher level compared to March and continues to increase. This is in agreement with news reports about the worse condition of Delhi now as compared to March \cite{debib}.

We obtained similar results for the states of Tamil Nadu, Gujarat and West Bengal which are the next worst-hit states in India.

\subsection{Kerala}
\begin{figure}[h]
    \centering
    \includegraphics[height=1.5in, width=8.0cm]{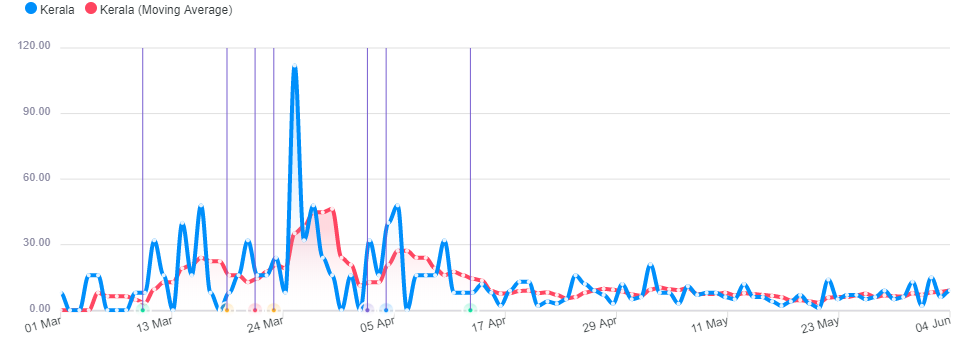}
    \caption{Twitter Word Count/Day Plot for Kerala}
    \label{fig:ker}
\end{figure}

Kerala provides an interesting counter-case study to the examples we have provided above. Kerala was the first state in India to identify a confirmed case of COVID-19 \cite{ker1}, and has tackled the situation well. It is observing a declining curve for the number of active cases, while the rest of the country continues to witness a surge in numbers. The state was able to ensure that the health facilities do not face shortage of critical equipment \cite{ker2}, and kept checks on overcrowding at hospitals \cite{ker4}. The state performed better compared to other states in the country like Maharashtra and Delhi and its model to combat the COVID-19 pandemic is being studied as a case study \cite{ker3}. Kerala has also reported a low death toll of only 14 deaths\footnote{https://www.mohfw.gov.in/},  which indicates that the health facilities weren’t burdened to the extent other states are suffering. This trend is reflected in our model's plot as the values have remained low since the beginning of the study, and has stagnated at a level close to 0 since 15th April 2020 (Fig. \ref{fig:ker}). The plot, corresponding tweets and news articles validate our claim that the model is successfully able to capture that the state has remained free of any incidences of overcrowding or shortage of critical equipment. 

\section{Conclusion}
From the literature review and results obtained, we can conclude that information obtained from Twitter data can provide useful insights about disease spread and its impact on the healthcare system. Twitter can provide trends about the ground reality of the burden on medical
facilities, which might not be captured in the official government reports. We found increasing signals and spikes, which were in accordance with the increase in the number of COVID-19 cases, as well as the incidences of overcrowding at hospitals as confirmed by the
news reports. Thus, researchers and epidemiologists can expand their range of methods used for monitoring of the COVID-19 pandemic by using the Twitter data model, as described in this paper. However, Twitter cannot provide all answers, and it may not be reliable for certain types of information. A significant limitation of the model is that social media is a platform where users can freely post anything, and thus, there is no way to verify the claims of any individual tweet. Therefore, we are relying on the assumption that if thousands of people are tweeting an incident, it is real and worth reporting. However, these need to be verified by trustable sources such as verified news articles to establish the claims reported by the twitter data.

\bibliographystyle{named}
\bibliography{main}

\end{document}